\documentstyle[12pt]{article}
\def\be{\begin{equation}}
\def\ee{\end{equation}}
\def\ba{\begin{eqnarray}}
\def\ea{\end{eqnarray}}
\def\ga{\mathrel{\raise.3ex\hbox{$>$\kern-.75em\lower1ex\hbox{$\sim$}}}}
\def\la{\mathrel{\raise.3ex\hbox{$<$\kern-.75em\lower1ex\hbox{$\sim$}}}}

\newcommand{\bi}[1]{\bibitem{#1}}
\newcommand{\fr}[2]{\frac{#1}{#2}}

\begin{document}
\baselineskip=16pt
\begin{titlepage} 
\rightline{UMN--TH--1820/99}
\rightline{TPI--MINN--99/40}
\rightline{OUTP-99-48P} 
\rightline{hep-ph/9909481}
\rightline{September 1999}  
\begin{center}

\vspace{0.5cm}

\large {\bf Cosmological 3-Brane Solutions}
\vspace*{5mm}
\normalsize

{\bf Panagiota Kanti$^1$, Ian I. Kogan$^2$, Keith A. Olive$^1$} and {\bf
Maxim  Pospelov$^1$}

\smallskip 
\medskip 
 
$^1${\it Theoretical Physics Institute, School of Physics and
Astronomy,\\  University of Minnesota, Minneapolis, MN 55455, USA} 

$^2${\it Theoretical Physics, Department of Physics, Oxford University}\\
{\it 1 Keble Road, Oxford, OX1 3NP,  UK}
\smallskip 
\end{center} 
\vskip0.6in 
 
\centerline{\large\bf Abstract}
\vspace*{2mm} 
%\smallskip\newline
We analyze cosmological equations in the brane world scenario with
one extra space-like dimension. We demonstrate that the cosmological 
equations can be reduced to the usual 4D Friedmann type if   
the bulk energy-momentum tensor is different from zero. 
We then generalize these equations to the case of a brane of finite 
thickness. We also demonstrate that when the bulk energy-momentum tensor is
different from zero, the extra space-like dimension can be compactified
with a single brane and show that the stability of the radius of
compactification implies standard cosmology and vice versa. For a brane of
finite thickness, we provide a solution such that the 4D Planck scale is
related to the fundamental scale by the thickness of the brane. In this
case, compactification of the extra dimension is unnecessary.

\end{titlepage} 

\section{Introduction} 

Motivated by the discovery of M-theory \cite{w,m}, there has been a
tremendous increase of interest in a class of models
(scenarios) with gravity and observable matter placed in a different
number of space-like dimensions. This idea presents us with the enticing
possibility to explain some long-standing particle physics problems 
by geometrical means. In M-theory, the correspondence of different string
theories (through dualities) is achieved in an 11D framework. Generally,
from a cosmological point of view, the scale factor of the 11th dimension
can be related to the expectation value of the string dilaton. 
In the particular case of the dimensional reduction to heterotic string
theory, where the hidden and matter sectors lie on separate 10D branes,
the size of the compact 11th dimension is relatively large, $r_{11} >
M_P$, so that one can relate the fundamental Planck scale with that of the
GUT scale. 

Taking this idea one step further, 
there are now several scenarios which try to relate the electroweak scale 
and the masses of observed particles with the fundamental higher dimensional
Plank scale exploiting either large extra space-like 
dimensions \cite{large,D1,D2} or an exponential scaling of the 
``warp'' factor in extra dimensions \cite{RS}. We note that warp factors
 in a context of M-theory have been discussed earlier in \cite{w,warp}.
 Both approaches 
assume that 
the SM particles are confined to a 3+1-dimensional slice (``brane'') 
of the $n$+3+1-dimensional space-time. Gravity is assumed to exist in the 
full higher dimensional space, ``bulk''. 
Related static domain-wall solutions in $N=1$ supergravity were
considered in \cite{cvetic}.
In the first proposal \cite{D1},
the fundamental gravitational scale $M_*$ 
can be related to the usual 4D Plank scale via a volume factor, $M_{\rm
Pl}^2=M_*^{n+2}r^n$, where $r^n$ is the volume of the compact space. If
$r$ is sufficiently  large,
$M_*$ can be as low as 1 TeV, thus providing a possible explanation to the
gauge hierarchy problem.  In the second proposal \cite{RS}, one posits
the configuration of a ``gravitational condenser''. Two branes of opposite
tensions, which gravitationally repel each other, are stabilized due to
the  negative cosmological constant in the bulk. As a result,  the distance
scales on the brane with negative tension are  exponentially smaller than
those on the positive tension brane, which can also explain why
$M_{\rm Pl}\gg M_W$. 

The possibility that there exists an extra dimension or dimensions allows
for a host of non-trivial phenomenology including the
production of Kaluza-Klein  excitations of gravitons at future colliders
or their detection in high-precision measurements at low energies.   
Moreover, they drastically change the cosmology of  early universe
\cite{cosm,LK}, sometimes too drastically to be consistent with  the
observable world. Thus, for example, a $1$  TeV-scale gravity scenario
excludes a universe hotter than 
$\sim$1 GeV due to an enormous emission of bulk modes at higher
temperatures \cite{D2}. 

Another serious problem emphasized recently is an unusual form of the 
Friedmann equations for the case of one extra dimension \cite{LK,LOW,BDL} 
which leads to a rather peculiar  
behavior of the Hubble parameter for the matter on the brane 
\be
H^2 =\left(\fr{\dot a}{a}\right)^2=\fr{\rho^2}{36M_*^6}.
\label{crazy}
\ee
This solution, subsequently checked in \cite{Berkl,Cline,Kor}, suggests 
that the gravitational law no longer reduces to Newton's law on the
3-brane. While mathematically correct as a solution to Einstein's
equations, this behavior, if nothing else, indicates that some key
ingredient is missing if this theory is to be capable of describing our
Universe. It is worth noting  that this unusual behavior of
$H$ does  not depend on the size of the extra dimension, but holds even for
$r\sim  M_*^{-1}\sim M_{\rm Pl}^{-1}$. 
If the solution (\ref{crazy}) is generic,
it poses a most serious threat to the viability of ``brane world'' 
scenarios. 

Therefore, it is important to review the assumptions under 
which the solution (\ref{crazy}) was obtained. To begin with, it is
generally assumed that the size of the extra
dimension is fixed in time. Since, as was noted earlier, this amounts
to fixing the vev of the dilaton, this is a reasonable assumption.
A rolling dilaton implies the variation of gauge couplings and particle
masses and, as with the Planck scale, there are very strong
constraints against this \cite{co}. In addition, there are several other
assumptions which go into the derivation of (\ref{crazy}), which we list
as:
\begin{enumerate}
\item The cosmological constants on the wall and in the bulk vanish,
ie. $\Lambda_w,\; \Lambda_b=0$.  
\item The brane is of zero thickness, $\Delta$=0.
\item The emptiness of the bulk, $T_{\mu\nu}^{bulk}=0$.
    
\end{enumerate}
While all of these assumptions are plausible, none are imperative. In
particular, it is quite reasonable to expect the brane to have
some finite thickness on the order of the fundamental scale or larger,
$\Delta>M_*^{-1}$. The emptiness of the bulk, $T_{\mu\nu}^{bulk}=0$,
cannot be a generic property either. Interactions between the brane and
gravity as well as other bulk mechanisms that might be responsible for 
the stabilization of radii should lead to a non-zero energy-momentum
tensor in the bulk. Finally, the condition on the vanishing of the 
cosmological constants can certainly be lifted as the brane can have its
own energy density, not related to the observable matter density
$\rho$. 

The case of nonvanishing $\Lambda_w$ and $\Lambda_b$ 
was studied by several groups \cite{K,Berkl,Cline,Kor}. 
The use of the two brane construction \cite{RS} allows one to obtain
the correct linear dependence of $H^2$ with $\rho$. Instead of $\rho^2$
in  (\ref{crazy}), one would have $(\Lambda_w+\rho)^2= \Lambda_w^2 + 
2\Lambda_w\rho+ \rho^2$ with $\Lambda_w^2$ term canceled 
by negative $\Lambda_b$.  As we will show, 
the introduction of two branes with $\pm \Lambda_w$ tensions and bulk
cosmological constant is not the unique way of recovering the
conventional  Friedmann equations.  

In this letter we carefully analyze Einstein's equations in 
4+1-dimensions with matter confined to a three-dimensional brane. In
particular,  we will derive sufficient conditions which ensure
a smooth transition to conventional cosmology and Newton's law on the
brane. To do this we relax the assumptions that $\Delta=0$ and 
$T_{\mu\nu}^{bulk}= 0$. 
If  $T_{\mu\nu}^{bulk} \neq 0$, the transition to the
conventional cosmology can be obtained with or without a vanishing
$\Delta$. We further demonstrate that the extra space-like dimension in
the thin-wall approximation can be compactified  for a single brane and
show that the stability of the radius of  compactification implies
standard cosmology and vice versa.  We then generalize this result to the
case of a finite wall thickness  and give a specific solution in this
case as well. Nevertheless, we find that finite $\Delta$ and vanishing bulk
energy-momentum tensor still lead to a phenomenologically  unacceptable
solution. 

\vspace*{10mm}

\section{The Theoretical Framework}

We start our analysis by considering the following 5-dimensional
theory describing the coupling of the matter content of the
universe with gravity
%%%%%%%%%%
\be
S=\int d^5x \sqrt{-\hat{g}}\,\Bigl\{\frac{M_5^3}{16\pi}\,\hat{R}
+ \hat{\cal L}_o \Bigr\}\,,
\label{action}
\ee
%%%%%%%%%%%
where $M_5$ is the fundamental five dimensional Planck mass and the hat
denotes 5-dimensio\-nal quantities. $\hat{\cal L}_o$ represents all other
contributions to the action which are not strictly gravi\-tational.  These
include the brane itself, matter on the brane, as well as any interaction
between the brane and the bulk. As 5D Poincar\'e invariance is 
broken by the brane, we require only that $\hat{\cal L}_o$  respect the
surviving 4D Poincar\'e invariance. We also consider the following ansatz
for the line-element of the 5-dimensional manifold
%%%%%%%%%
\be
ds^2=-n^2(t,y) dt^2 + a^2(t,y) \delta_{ij} dx^i dx^j + b^2(t,y) dy^2\,,
\label{metric}
\ee
%%%%%%%%%
where $\{t,x^i\}$ and $y$ denote the 4-dimensional spacetime (in the
direction of the brane) and the extra dimension, respectively.

The variation of the action functional (\ref{action}) with respect to
the 5-dimensional metric tensor $\hat{g}_{MN}$ leads to the Einstein's
equations which for the above spacetime background take the form
(see e.g. \cite{LOW, BDL})
%%%%%%%%%%%%%
\vskip .005in
\ba
\hat{G}_{00} &=& 3\Biggl\{\frac{\dot{a}}{a}\,\Biggl(\frac{\dot{a}}{a} +
\frac{\dot{b}}{b}\Biggr) -\frac{n^2}{b^2}\,\Biggl[\frac{a''}{a} +
\frac{a'}{a}\,\Biggl(\frac{a'}{a} - \frac{b'}{b}\Biggr)\Biggr]\Biggr\}
= \hat{\kappa}^2 \, \hat{T}_{00}\,,\label{00}\\[4mm] 
\hat{G}_{ii} &=& \frac{a^2}{b^2}\Biggl\{\frac{a'}{a}\,
\Biggl(\frac{a'}{a} + 2\frac{n'}{n}\Biggr) -\frac{b'}{b}\,\Biggl(\frac{n'}{n}
+2\frac{a'}{a}\Biggr) +2 \frac{a''}{a} +\frac{n''}{n}\Biggr\}\nonumber\\[4mm]
&+& \frac{a^2}{n^2}\Biggl\{\frac{\dot{a}}{a}\,\Biggl(-\frac{\dot{a}}{a} +
2\frac{\dot{n}}{n}\Biggr) -2\frac{\ddot{a}}{a}+ \frac{\dot{b}}{b}\,
\Biggl(-2\frac{\dot{a}}{a} + \frac{\dot{n}}{n}\Biggr) -
\frac{\ddot{b}}{b}\Biggr\}= \hat{\kappa}^2\,\hat{T}_{ii}\,,
\label{ii}\\[4mm]
\hat{G}_{05} &=& 3\Biggl(\frac{n'}{n} \frac{\dot{a}}{a}
+ \frac{a'}{a} \frac{\dot{b}}{b} -\frac{\dot{a}'}{a}\Biggr)=0\,,
\label{05}\\[4mm]
\hat{G}_{55} &=& 3\Biggl\{\frac{a'}{a}\,\Biggl(\frac{a'}{a} +
\frac{n'}{n}\Biggr) -\frac{b^2}{n^2}\,\Biggl[\frac{\dot{a}}{a}\,
\Biggl(\frac{\dot{a}}{a}-\frac{\dot{n}}{n}\Biggr) +
\frac{\ddot{a}}{a}\Biggr]\Biggr\} = \hat{\kappa}^2\,\hat{T}_{55}\,, 
\label{55}
\ea
%%%%%%%%%%%%%
where $\hat{\kappa}^2=8\pi \hat{G}=8\pi/M_5^3$. Note that, in the above
relations, the dots and primes denote differentiation with respect to
$t$ and $y$, respectively. 

In this paper, we assume that the scale factor of
the fifth dimension depends neither on space nor time, i.e.
$b=const$. In that case, the $(05)$-component of Einstein's equations can
be integrated to give the result
%%%%%%%%%%%
\be
n(t,y)=\lambda(t)\,\dot{a}(t,y)\,.
\label{soln}
\ee
%%%%%%%%%%
while the $(00)$-component reduces to a differential equation
for $a$ with respect to $y$ with the general solution depending on
the form of the energy density $\hat{\rho}$ of the universe. For
future reference, note 
that, by choosing the normalization $n(t,y=0)=1$, the Hubble
parameter can be expressed in terms of $\lambda(t)$ in the following way
%%%%%%%%%%%
\be
H^2 \equiv \biggl(\frac{\dot{a}_0}{a_0}\biggr)^2=
\frac{1}{\lambda^2(t) a_0^2(t)}\,,
\label{hubble}
\ee
%%%%%%%%%%%%
where the subscript $0$ denotes quantities evaluated at $y=0$.
\paragraph{}
In \cite{BDL}, it was assumed that the usual matter content
of our universe is confined to a 4-dimensional hypersurface located
at $y=0$. In this case, the energy-momentum tensor of our brane-universe
can be expressed in the form 
%%%%%%%%%%%
\be
T^A_{\,\,\,\,\,B}=\frac{\delta(y)}{b}\,{\rm diag} (-\rho,
p, p, p, 0)
\ee
%%%%%%%%%%%%%
The inhomogeneity in the distribution of matter in the 5-dimensional
spacetime leads to the discontinuity of the first derivative of the
metric tensor with respect to $y$ and, thus, to the appearance of
a Dirac delta function in its second derivative. By matching the
coefficients of the delta functions that appear at both sides of
the $(00)$ and $(ii)$ components of Einstein's equations, the
{\it jumps} in the first derivatives of $a$ and $n$ were derived
and found to be
%%%%%%%%%%%%%
\ba
&~& [a']=a'(0^+)-a'(0^-)=-\frac{\hat{\kappa}^2}{3}\,\rho\,a_0 b_0\,,
\label{jumpa}\\[3mm]
&~& [n']=n'(0^+)-n'(0^-)=\frac{\hat{\kappa}^2}{3}\,(3p+2\rho)\,n_0 b_0\,.
\label{jumpn}
\ea
%%%%%%%%%%%%%%
For
cosmological solutions which are symmetric under the change 
$y \rightarrow -y$ and for an empty bulk-universe, the $(55)$ component
of Einstein's equations (\ref{55}) takes the form
%%%%%%%%%%%%%%
\be
\frac{\dot{a}_0^2}{a_0^2} + \frac{\ddot{a}_0}{a_0} =
-\frac{\hat{\kappa}^4}{36}\,\rho\,(\rho + 3p)\,.
\label{friedmann0}
\ee
%%%%%%%%%%%%%%
The above equation leads to
the 5-dimensional analog of the Friedmann equation given in eq.
(\ref{crazy}). However, as noted earlier, instead of the usual
$H^2 \propto \rho$ dependence, one finds $H \propto \rho$ implying a
departure from the standard cosmological expansion which would
follow from the Newtonian force law. In subsequent work \cite{paper2}, we
will show the precise effect on the gravitational force law which the
various brane solutions imply. 

In this class of exact cosmological
solutions  with 
$b=const$, the spatial scale factor $a$ decreases as one moves
away from our brane-universe at $y=0$. The global definition of this group
of {\it special} solutions throughout spacetime strongly relies on the
existence of a second brane at $y=1/2$ whose matter content is heavily
constrained by the energy density and pressure of the matter on our
brane. 

Here, we propose an alternative mechanism to restore
the usual form of the Friedmann equation, in the framework of a
5-dimensional gravitational theory, which does not necessitate the
introduction of a second brane or a cosmological constant. We will follow
two different approaches: we first 
 allow a non-vanishing energy-momentum
tensor in bulk and subsequently consider the case of a brane of finite 
thickness,
$\Delta$. The bulk energy-momentum tensor 
 is given by
%%%%%%%%%
\be
\hat{T}^A_B={\rm diag} (- \hat{\rho}, \hat{p}, \hat{p}, \hat{p},
\hat{T}^5_5)\,,
\ee
%%%%%%%%%
which, when combined with the equation for the conservation of the
energy-momentum tensor,
$D_M \hat{T}^M_{\,\,\,\,\,\,N}=0$ leads to the following relations 
%%%%%%%%%%%
\ba
\frac{d\hat{\rho}}{dt} + 3(\hat{\rho}+\hat{p})\,\frac{\dot{a}}{a} 
+ (\hat{\rho} + \hat{T}_5^5)\,\frac{\dot{b}}{b} &=& 0\,,
\label{zeroth} \\[5mm]
\Bigl(\hat{T}^5_5\Bigr)^{'} + \hat{T}^5_5\Biggl(\frac{n'}{n} +
3\frac{a'}{a}\Biggr)
+ \frac{n'}{n}\,\hat{\rho} -3 \frac{a'}{a}\,\hat{p} &=& 0\,.
\label{fifth}
\ea
%%%%%%%%%%%%
In the following sections, we will derive the explicit form for the
bulk energy-momentum tensor and
the corresponding metric, which is consistent with the above energy
conservation equations as well as the gravitational equations of motion
(\ref{00} -\ref{55}).

\section{Thin Wall Approximation}

In our first approach, we adopt the concept of a brane-universe with zero
thickness while allowing for a non-zero value of $\hat{T}^5_5$ in
the bulk.  To do so, we assume that the
energy momentum tensor on the brane has the form
%%%%%%%%%
\be
\hat{T}^A_B={\rm diag} \Bigl(\frac{\delta(y)}{b}(-\rho, p, p, p),
T^5_5\Bigr)\,,
\ee
%%%%%%%%%
while outside the brane, $\hat{\rho}=\hat{p}=0$, but $\hat{T}^5_5$ 
retains a non-zero value consistent with the energy conservation
equations. For a constant scale factor along the extra dimension, the
$(00)$-component of Einstein's equations can be easily integrated and
gives the following general solution for $a$ 
%%%%%%%%%
\be
a^2(t,y)= a_0^2(t) + c(t)\,|y| + \frac{b^2}{\lambda^2(t)}\,y^2 \,,
\label{gen}
\ee
%%%%%%%%%
outside the brane.
The unknown function $c(t)$ can be determined from the {\it jump} of the
first derivative of $a$, given by eq.(\ref{jumpa}), and is found to be
%%%%%%%%%%
\be
c(t)= -\frac{\hat{\kappa}^2}{3}\,\rho\,a_0^2\,b\,.
\ee
%%%%%%%%%%

The general solution (\ref{gen}) for $a^2$ has always a minimum at
%%%%%%%%%%%
\be 
y_{min}(t)=\pm\,\frac{c(t) \lambda^2(t)}{2b^2}= \pm\,\frac{\hat{\kappa}^2 \rho}
{6b\,(\dot{a}_0/a_0)^2}
\ee
%%%%%%%%%%%
which in turn, by making use of eq.~(\ref{hubble}), leads to the
evolution equation for $a_0$
%%%%%%%%%%%%
\be
\biggl(\frac{\dot{a}_0}{a_0}\biggr)^2 = 
\frac{\hat{\kappa}^2 {\rho}}{3\,(2b|y_{min}(t)|)}\,.
\label{friedmann}
\ee
%%%%%%%%%%%%
From the above equation, one can immediately conclude that the Friedmann
equation with the correct linear dependence on the density $\rho$
is recovered if, and only if, $y_{min}(t)=const$. In that case, we can
identify
%%%%%%%%%%
\be
\kappa^2=\frac{\hat{\kappa}^2}{2b|y_{min}|} \,\,\Rightarrow\,\,
M_P^2=M_5^3\,(2b|y_{min}|)\,.
\label{kappa}
\ee
%%%%%%%%%%%

As we will, now, demonstrate, the time-independence of $y_{min}$ and,
thus, the restoration of the Friedmann equation strongly relies on a
non-vanishing $\hat{T}^5_5$ in the bulk. Indeed, by
substituting the solution (\ref{gen}) for $a$ in the $(55)$-component
of Einstein's equations, we obtain the constraint
%%%%%%%%%%%%%%
\be
c(t)\,\dot{c}(t)-\frac{4b^2}{\lambda^2(t)} \dot{a}_0(t) a_0(t) +
\frac{4b^2}{\lambda^3(t)} \dot{\lambda}(t) a_0^2(t) = 
\frac{\hat{\kappa}^2}{3}\,4b^2 a^3(t,y) \dot{a}(t,y)\,\hat{T}^5_5\,,
\label{con}
\ee
%%%%%%%%%%%%%%%
which, when integrated with respect to time, leads to the following
expression
%%%%%%%%%%%%%
\be
\biggl(\frac{\dot{a}_0}{a_0}\biggr)^2 \equiv \frac{1}{\lambda^2 a_0^2}=
\frac{\hat{\kappa}^4 \rho^2}{36} - \frac{2 \hat{\kappa}^2}{3 a_0^4}
\int a^3(t,y)\,\dot{a}(t,y)\,\hat{T}^5_5\,dt\,\,,
\label{general}
\ee
%%%%%%%%%%%%
for the Friedmann equation. From the above result, we can easily see that,
if we choose $\hat{T}_5^5=0$, we recover the behavior $H \sim \rho$,
which is characteristic of the case of an empty bulk-universe~\cite{BDL}.
Clearly, a non-vanishing value for $\hat{T}_5^5$ is required
to recover the standard Friedmann expansion.

Any expression for $\hat{T}_5^5 \ne 0$ must be consistent with the
general conservation of $\hat{T}^M_{\,\,\,\,N}$   and in
particular the fifth component of those equations (\ref{fifth}).
For vanishing $\hat{\rho}$ and $\hat{p}$ in the bulk, the general
solution of the aforementioned component takes the form
%%%%%%%%%%%%
\be
\hat{T}^5_5=\frac{ w(t)}{n(t,y)\,a^3(t,y)}\,.
\label{gent}
\ee
%%%%%%%%%%%%%%
The above expression allows us to make
a suitable choice for $\hat{T}^5_5$ that will cancel the quadratic
dependence on $\rho$ in (\ref{general}). Indeed, we find that the
following non-vanishing bulk value of $\hat{T}^5_5$
%%%%%%%%%%%%%
\be
\hat{T}_5^5=-\frac{a_0^3(t)}{2 n(t,y) a^3(t,y)}\,\Biggl[\frac{(\rho-3p)}
{2b|y_{min}|}
+ \frac{\hat{\kappa}^2}{6}\,\rho\,(\rho+3p)\Biggr]
\label{ansatz}
\ee
%%%%%%%%%%%%%% 
leads to the time-independence of $y_{min}$ and to the restoration of
the Friedmann equation (\ref{friedmann}), at the same time. Given the fact
that $\rho$ and $p$, being defined at the origin, are
functions of time only, the ansatz (\ref{ansatz}) for
$\hat{T}^5_5$ obviously belongs to the class of solutions given by
(\ref{gent}).

In eqs.~(\ref{kappa}) and (\ref{ansatz}), $2by_{min}$ represents the
length scale that determines the 4-dimensio\-nal Planck mass $M_P$
and the physical distance over which $\hat{T}^5_5$ is smoothly distributed
in the bulk. Moreover, this length can have another important interpretation.
It can be shown that both $a^2$ and $n^2$ reach their extrema at the
same point $y=y_{min}$. If one identifies the points 
$y=|y_{min}|$ and $y=-|y_{min}|$, the extra dimension is effectively
compactified with the size of the compact dimension being given by
$2by_{min}$.

Before concluding this section, we would like to stress the importance
of the bulk energy-momentum tensor. While one can certainly construct
configurations (D-branes, solitons in some exotic field theory, etc.)
where there is no bulk energy-momentum tensor (or cosmological constant in
the bulk), in the thin wall case, there is no newtonian limit of
gravity and such a model can be safely discarded at once. Instead, we have
shown that one {\it must} have a non-vanishing bulk energy-momentum tensor
or require at least two branes with opposite cosmological constants in
addition to a bulk cosmological constant.

% First of all we see that because $a(y,t)$ grows with $y$ the stress
%energy tensor 
% components are decaying as $1/a^3$. If scale factor $a$ grows
%exponetially (like in 
% \cite{RS}) than we simply have exponential decay of energy-momentum
%tensor far  from the brane - the natural law for distribution of energy
%near any soliton. In case of a power decay  the cituation is different,
%but here we are dealing  with purely gravitational effect, when  due to
%long-range nature of gravitational field the original distribution of
%energy in a flat space becomes less localized and there are power, rather
%than exponetial "tails".

\section{Thick Wall Approximation}

In the second approach, we assume that our brane-universe has a
non-vanishing thickness $2\Delta$ and the energy density $\hat{\rho}$
is homogeneously distributed over the 5-dimensional spacetime of our
brane. For zero pressure on the brane, the zeroth component
(\ref{zeroth}) of the equation for the conservation of energy gives 
$\hat{\rho}=\hat{\rho}_0/a^3$, where $\hat{\rho}_0$ is a constant
both in $t$ and $y$. In that case, the general solution for the spatial
scale factor $a$, inside the brane, takes the form
%%%%%%%%%%%%%
\ba
&~& \hspace*{-1.5cm}\frac{2A^2}{B^3(t)}\,\log\Biggl(\frac{2}{B(t)}\,
\Bigl[B^2(t) a(t,y)-A^2\Bigr] +
2 \sqrt{E_{in}(t) + B^2(t) a^2(t,y) -2 A^2 a(t,y)} \Biggr) \nonumber\\[4mm]
&~& \hspace*{1.5cm} +\, \frac{2}{B^2(t)}\, \sqrt{E_{in}(t) +
B^2(t) a^2(t,y) -2 A^2 a(t,y)}=
\pm \sqrt{2}\,[|y|+ C_{in}(t)]\,,
\label{sola}
\ea
%%%%%%%%%%%%%%%
where
%%%%%%%%%%%%
\be
A^2= \frac{2b^2 \hat{\kappa}^2 \hat{\rho}_0}{3}\,,
\qquad B^2(t)=\frac{2b^2}{\lambda^2(t)}
\ee
%%%%%%%%%%%
and $E_{in}(t)$ and $C_{in}(t)$ are unknown functions of time which need
to be determined. The $\pm$ sign at the r.h.s. of eq.~(\ref{sola})
corresponds to the sign of the first derivative of $a$ with respect
to $y$, inside the brane, which causes the scale factor to increase
or decrease, respectively, as we move away from the origin. 
The symmetry of our cosmological solution, exhibited when $y \rightarrow -y$,
leads to the vanishing of $a'$ at the origin or equivalently to the
condition
%%%%%%%%%%%%
\be
E_{in}(t)= 2A^2 a_0(t)-B^2(t) a^2_0(t)\,.
\label{dazero}
\ee
%%%%%%%%%%%%%
The function $C_{in}(t)$ can be determined by evaluating the 
solution (\ref{sola}) at $y=0$, and is found to be
%%%%%%%%%%
\be
C_{in}(t)=\pm \frac{\sqrt{2}A^2}{B^3(t)}\,
\log\Biggl(\frac{2}{B(t)}\,\Bigl[B^2(t) a_0(t)-A^2\Bigr]\Biggr)\,.
\ee
%%%%%%%%%%

In order to determine $E_{in}(t)$, we substitute the implicit solution
for $a$ (\ref{sola}) in the remaining components of Einstein's equations.
While the $(ii)$ component is trivially satisfied, the $(55)$ component
leads to the following constraint
%%%%%%%%%%
\be
\frac{d E_{in}(t)}{dt}=\frac{4b^2}{3}\hat{\kappa}^2\,a^3(t,y) \dot{a}(t,y)\,
(\hat{\rho} + \hat{T}^5_5)\,.
\label{conein}
\ee
%%%%%%%%%%%
The fifth component of the conservation of the energy-momentum tensor
$\hat{T}^M_{\,\,\,\,N}$ (\ref{fifth}) imposes the existence of a
non-vanishing $\hat{T}^5_5$ on the brane. Moreover, the choice
$\hat{T}^5_5 = 0$ would lead to an inconsistency in the above equation
since $E_{in}$ is, by definition, a function of time only. 

The results of the previous section show that a non-vanishing bulk
$\hat{T}^5_5$ gives us the extra degree of freedom to recover the Friedmann
equation on the brane.  For the thick wall solution, we will demonstrate
that, once again, the appropriate choice of $\hat{T}^5_5$ lead to the
usual dependence of the Hubble parameter on the energy density. As
before, although the usual matter content should be localized on the
brane-universe, the fifth component of the energy-momentum tensor
$\hat{T}^5_5$ must be smoothly distributed over the entire extra
dimension. From the condition
(\ref{dazero}), we can easily see that the usual form of the Friedmann
equation can be obtained for $E_{in}=0$. This is consistent with
eq.~(\ref{conein}) only if $\hat{T}^5_5$ is exactly equal and opposite
to the energy density $\hat{\rho}$ of our brane-universe. In that case,
we obtain
%%%%%%%%%%%
\be
\Biggl(\frac{\dot{a}_0}{a_0}\Biggr)^2=\frac{\kappa^2 \rho(t,0)}{3}\,,
\label{newton2}
\ee
%%%%%%%%%%%%
where we have used the fact that $\hat{\rho}=\rho/(2\Delta b)$ and defined
%%%%%%%%%%%%
\be
\kappa^2=\frac{\hat{\kappa}^2}{\Delta b} \,\,\Rightarrow\,\,
M_P^2=M_5^3\,(\Delta b)\,.
\label{def}
\ee
Note that the
above expression for the Friedmann equation involves the 4-dimensional
energy density $\rho$ and gravitational constant $\kappa^2$, in
distinction to the result (\ref{friedmann0}) where the square of the
4-dimensional energy density was combined with the 5-dimensional
gravitational constant $\hat{\kappa}^2$. Here, we want to note that the
above choice, $E_{in}=0$, corresponds only to a special solution of the
system and that other solutions also leading to the restoration of the
Friedmann equation do exist and will be discussed elsewhere~\cite{paper2}.

By making use of the condition (\ref{dazero}), for the choice $E_{in}=0$,
the implicit solution (\ref{sola}) for $a$ inside the brane takes the
form
%%%%%%%%%%%%
\ba
&~& \hspace*{-1cm}\frac{a_0(t)}{B(t)} \log\Biggl(B(t)
[2a(t,y)-a_0(t)] + 2 \sqrt{B^2(t) a(t,y)[a(t,y)-a_0(t)\bigr]}
\Biggr)\nonumber\\[4mm]
&~& \hspace*{2cm} +\,2\sqrt{\frac{a(t,y)}{B^2(t)} [a(t,y)-a_0(t)]}=
\sqrt{2}\,\Biggl(y+ \frac{a_0(t)}{2B(t)} \log\Bigl[B(t) a_0(t)\Bigr]\Biggr)\,,
\label{sola2}
\ea
%%%%%%%%%%%%%%%
from which it follows that $a(t,y) \geq a_0(t)$ leading to the conclusion
that the spatial derivative of the scale factor $a$, inside the brane,
should be positive and, thus, $a$ increases as we move away from the origin.
\paragraph{}
In order to complete the picture, we need to determine the solution for
the scale factor $a$ outside the brane. For vanishing $\hat{\rho}$,
the general solution for $a$ can be written as
%%%%%%%%%%%%%
\be
a^2(t,y)=\frac{b^2}{\lambda^2(t)}\,[\,|y|+C_{out}(t)]^2 -
\frac{\lambda^2(t) E_{out}(t)}{2b^2}\,,
\label{solout}
\ee
%%%%%%%%%%%%%%
where again $C_{out}(t)$ and $E_{out}(t)$ are functions of time which
need to be determined. If we substitute the above solution for $a$ in the
$(55)$-component of Einstein's equations, we obtain a constraint for
$E_{out}(t)$ similar to eq.~(\ref{conein}) but with $\hat{\rho}=0$, while
the $(ii)$-component is, again, trivially satisfied. By continuity, the
value of $\hat{T}^5_5$ outside the brane is given by
%%%%%%%%%%%%%%%
\be
\hat{T}^5_5= - \frac{\hat{\rho}_0\,n(t,\Delta)}{n(t,y)\,a^3(t,y)}\,.
\label{5out}
\ee
%%%%%%%%%%%%%%
We can easily see that the above expression reduces to the inside value of
$\hat{T}^5_5$ in the limit $y \rightarrow \Delta$ and is consistent with
the general solution (\ref{general}). The above expression,
when substituted in the $(55)$-component, leads to the
determination of $E_{out}(t)$ while $C_{out}(t)$ can be found by
evaluating the solution (\ref{solout}) at $y=\Delta$. We may, then,
write the final solution for the spatial scale factor outside the brane
in the following way
%%%%%%%%%%%%%
\be
a^2(t,y)=a_0(t) a(t,\Delta) + \frac{B^2(t)}{2}\,\Biggl[\,|y|-\Delta
+\sqrt{\frac{2a(t,\Delta)}{B^2(t)}\,[a(t,\Delta)-a_0(t)]}\,\Biggr]^2\,. 
\label{fsolout}
\ee
%%%%%%%%%%%%%%
Here, we have to note that the solutions derived for $a$, and $n$
through eq.(\ref{soln}), as well as their first derivatives with
respect to $y$, are continuous across the wall, i.e. at $y=\Delta$,
ensuring the smooth transition in spacetime from our matter dominated
brane-universe to a $\hat{T}^5_5$ dominated bulk-universe. 

Finally, let
us point out that the continuity of $a'$ across the wall forces the
spatial scale factor to increase as we move away from the
brane in this particular solution. (There are of course other thick
wall solutions where $a$ decreases with $y$, as in the thin-wall
solution.  We will discuss these solutions elsewhere \cite{paper2}.) The
absence of any minima for
$a$ outside the brane will cause the scale factor to increase indefinitely
for a non-compact extra dimension. However, this is not a problem since
the length scale that determines the 4-dimensional Planck scale $M_P$, in
terms of $M_5$, is the thickness $\Delta$ of the brane and not the size
of the extra dimension.

It is instructive to analyze the last remaining possibility of the wall 
with a finite thickness and the vanishing $\hat{T}^5_5$ in the bulk. 
As we noted earlier, directly from the 5th component of the energy conservation equation, 
we see that in the wall,  $\rho \ne 0$ implies that $\hat{T}^5_5 \ne 0$.
If we assume $\hat{T}^5_5 = 0$ in the bulk, we can use the continuity of $\hat{T}^5_5$ across
the wall and eq. (\ref{fifth}) to find the solution for $\hat{T}^5_5$ inside the wall:
%%%%%%%%%%%
\be
\hat{T}^5_5=\hat \rho\left[{n(t,\Delta)\over n(t,y)}-1\right ].
\label{t55inside}
\ee
%%%%%%%%%%%%
As before, we take $\hat p=0$. 
Using this solution and the $(55)$-component of Einstein's equations,
we obtain the evolution equation for $a_0$:
%%%%%%%%%%%%
\be
\Biggl(\frac{\dot{a}_0}{a_0}\Biggr)^2 +
\frac{\ddot{a}_0}{a_0} =
  -\frac{\hat{\kappa}^2}{6b}\,\rho\,\frac{n(t,\Delta)-1}{\Delta}
\label{diff}
\ee
%%%%%%%%%%%%
It is easy to see that the r.h.s of eq. (\ref{diff}) does not
correspond to the usual form of the Friedmann equation. 
Thus, we conclude that the case of $\Delta \neq 0$
and $\hat{T}^5_5=0$ in the bulk does not lead to the standard
cosmological expansion for the spatial scale factor.

%Let us also note that in this case it is simply impossible to have 
%$\hat{T}^5_5=0$, this will contradict our major assumption that $b$
%is constant.  In some sense we derived here a no-go theorem - any solution
%with constant $b$ and finite thickness of a brane must have a bulk 
%$\hat{T}^5_5=0$ ! 

\section{Conclusions}

The idea that our four-dimensional spacetime is a slice of a higher
dimensional space time has always been an intriguing one. In most cases,
one generally assumes that the physical size of the extra dimension is
small (of order $M_P^{-1}$) and we can decompose the higher-dimensional
states in a Fourier expansion of momentum modes in the extra dimensions.
This is the usual Kaluza-Klein decomposition.  Of course there is no
inherent reason
that the size of the extra dimensions need be small (or even compact!). 
Our Universe as a three-brane however, is not a trivial proposition. As
shown in several recent works, the cosmological solutions to Einstein's
equations for a three-brane embedded in a higher dimensional space-time,
leads to the unphysical solution that the Hubble parameter on the 3-brane
is proportional to the matter density.  This expansion law is not
consistent with Newtonian gravity. This conundrum can be fixed by
balancing the cosmological constants on the brane and in the bulk but
requires two branes for a complete solution.

In this paper, we have derived alternative solutions for recovering 
the normal Hubble expansion on the three-brane.  Our
solutions do not require the existence of a second brane.  In both types
of solutions we have presented, we require that $\hat{T}^5_5$ is
non-vanishing in the bulk.  Keeping to the thin wall approximation, we
have shown that by an appropriate choice of $\hat{T}^5_5$, the 3-space
scale factor which decreases as we move away from the brane, has a fixed
minimum.  As such, by identifying the points $\pm y_{min}$, we can
compactify the extra dimension without the need of a second brane.
The 4D Plank scale is then determined
by the compactification radius, $M_P^2 = M_5^3 (2by_{min})$. In the thick
wall solution, the size of the extra dimension remains infinite, and the
Planck scale is determined by the thickness of the wall, $M_P^2 = M_5^3
(2b\Delta)$.

We also want to stress that in any realization of  
the world as a brane scenario
 in which there is a 4-dimensional  Einstein gravity 
on a brane, one must have the normal
 law $H^2 \sim \rho$ which simply follows from newtonian 
limit of general relativity. To do so, it appears that the bulk can not be ignored.
Either there must exist a bulk cosmological constant or a non-vanishing 
$\hat{T}^5_5$.

In this paper, we have not specified any particular source which accounts
for a non-vanishing $\hat{T}^5_5$. This alone is worthy of a separate
investigation. However, we note that the normal 4D form of the Friedmann
equations can be recovered if the value of $\hat{T}^5_5$ on the brane
is proportional to the trace of the 4D energy-momentum tensor.
This bears a strong resemblance to the case of the dilaton interaction
with matter, which is also proportional to $T_\mu^\mu$ on the brane.
However, we cannot claim that this solution is unique. In the case of the
thin wall, where the extra dimension is compact, the existence of
$\hat{T}^5_5$ in the bulk should be related to the physics responsible
for the  stabilization of the extra dimension, and perhaps equivalently
the stabilization of the dilaton vev. These and other related issues
will be addressed in a future publication \cite{paper2}.

This work was supported in part by the Department of Energy
under Grant No.\ DE-FG-02-94-ER-40823 at the University of Minnesota, and
PPARC grant GR/L56565 and INTAS 95-RFBR-567 at Oxford. 
One of us (I.K.) wants to thank TPI and Department of Physics, 
University of Minnesota for its hospitality during the summer
of 1999 where this work began.

%%%%%%%%

\end{document}